\documentclass[12pt]{article}

\usepackage{scicite}

\usepackage{times}
\usepackage{graphicx}
\usepackage{setspace}

\newenvironment{sciabstract}{%
\begin{quote} \bf}
{\end{quote}}

\newcounter{lastnote}
\newenvironment{scilastnote}{%
\setcounter{lastnote}{\value{enumiv}}%
\addtocounter{lastnote}{+1}%
\begin{list}%
{\arabic{lastnote}.}
{\setlength{\leftmargin}{.22in}}
{\setlength{\labelsep}{.5em}}}
{\end{list}}

\title{From Gas to Stars Over Cosmic Time}

\author
{Mordecai-Mark Mac Low$^{1,2\ast}$\\
\\
\normalsize{$^{1}$Department of Astrophysics, American Museum of
  Natural History}\\
\normalsize{79th Street at Central Park West, New York, NY 10024, USA}\\
\normalsize{$^{2}$Zentrum der Astrophysik der Universit\"at
  Heidelberg, Institut f\"ur Theoretische Astrophysik} \\
\normalsize{Albert-Ueberle-Str.\ 2, 69121 Heidelberg, Germany}
 \\
\\
\normalsize{$^\ast$To whom correspondence should be addressed; E-mail:  mordecai@amnh.org.}
}

\date{}

\begin{document} 


\baselineskip24pt


\maketitle 

\singlespacing


\begin{sciabstract} 
  From the time the first stars formed over 13 billion years ago to
  the present, star formation has had an unexpectedly dynamic history.
  At first, the star formation rate density increased dramatically,
  reaching a peak 10 billion years ago more than ten times the present
  day value.  Observations of the initial rise in star formation
  remain difficult, poorly constraining it. Theoretical modeling has trouble predicting this
  history because of the difficulty in following the feedback of
  energy from stellar radiation and supernova explosions into the gas
  from which further stars form.  Observations from the ground and
  space with the next generation of instruments should reveal the full
  history of star formation in the universe, while simulations appear
  poised to accurately predict the observed history.
\end{sciabstract}


Once, there were no stars.  This simple, yet profound, statement is an
inevitable consequence of Big Bang cosmology.  Observations of the
cosmic microwave background reveal that at the time of its emission
some 300,000 years after the Big Bang, the Universe was filled with
hydrogen and helium having a density uniform to a few parts in 10$^5$
\cite{smoot1992,spergel2007}. Understanding how that gas evolved into
the Universe filled with stars that we observe today, $13.798 \pm 0.037$
billion years later \cite{jarosik2011,planck-collaboration2013}, remains one of the most important
goals of modern astrophysics.

The ordinary matter from which stars form makes up only $4.63 \pm 0.0024$\% of the
total mass-energy of the Universe and 17\% of the matter
\cite{bennett2012}, with the rest being made up of not yet identified
cold dark matter, and dark energy.  Star formation traces the
gravitational collapse of the dark matter into the cosmic web in a
universe whose expansion is currently dominated by dark energy.

\framebox{\begin{minipage}{6in}\paragraph{Star formation} Stars form when gravity causes interstellar gas and dust---the
interstellar medium---to collapse towards
regions of higher density, while radiative cooling prevents the
temperature from increasing as densities get higher, so that pressure
cannot prevent collapse.  Collapse ultimately continues until central
temperatures and pressures rise high enough for nuclear fusion to
begin, heating the gas sufficiently to counterbalance the force of
gravity and maintain hydrostatic equilibrium.  This process occurs
over a time of order the free-fall time of the gas $t_{\rm ff} \sim
(G\rho)^{-1/2}$, depending only on  $\rho$, the mass density of
the gas, where  $G$ is the gravitational constant.  The mass of a region
capable of gravitational collapse is given by the Jeans
\cite{jeans1902} mass
$$
M_J  = (\pi / 6) G^{-3/2} \rho^{-1/2} c_s^3,
$$
where $c_s$ is the sound speed of the gas, which depends on the
the temperature as $T^{1/2}$.\end{minipage}}

The collapse of dark matter from small perturbations into the
gravitationally bound structures first identified as halos around
galaxies can be followed by the Press-Schechter
\cite{press1974,bond1991} analytic excursion-set 
formalism.  However, the properties of the
stellar populations that form within those halos, in visible galaxies,
depend critically on nonlinear physics including magnetized gas dynamics,
radiative transfer, and nuclear fusion.  We are ultimately forced to
rely on numerical simulations of galaxy formation 
incorporating approximate treatments of these processes in order to
predict the observable outcomes of cosmological theory.

The approximations used in these models rely on a detailed
understanding of the smaller-scale physics determining star
formation.  Feedback of energy into the interstellar and
intergalactic gas from ionizing ultraviolet (UV) radiation and supernova explosions
from stars, and from high-velocity outflows and radiation from supermassive
black holes, has been extremely difficult to model
well enough to predict the star formation history of galaxies.  Models that
neglect feedback or understate its importance predict far too much
star formation at early times \cite{white1991,springel2003}.
Because of its central importance to modern models of star formation,
I explore this topic in some detail in this review.
 
\section*{Star Formation History of the Universe}

Star formation in the Universe can be described using the star
formation rate (SFR) density at any time, often expressed in units of
solar masses per cubic megaparsec per year.  After a slow start, the
SFR density in the observed Universe peaked some 10 billion years ago,
when stars formed an order of magnitude faster than they do in the
present epoch \cite{madau1996,lilly1996,hopkins2006}.  Light emitted
then has been redshifted by the expansion of the Universe to longer
wavelengths by a factor of $(1+z) = 3$, so we refer to that
era\footnote{The exact relationship between redshift and lookback time
  is nonlinear and depends on the actual expansion law of the
  Universe.} as being at redshift $z=2$.  The time dependence of the
SFR at higher redshifts (greater lookback times) remains uncertain,
leaving the time of formation of the first stars poorly constrained
\cite{hopkins2006,bromm2011,trenti2012}. Two major lines of evidence,
from imaging galaxies, and gamma ray bursts, give apparently
conflicting results \cite{kistler2009,robertson2012,trenti2012}, that
in turn must agree with two constraints: the range of time over which
the intergalactic medium was reionized by UV radiation from stars, and
the well-observed density of stars at $z = 4$.

The first line of evidence relies on the detection of high redshift
galaxies by photometry in multiple colors.  Even broad band photometry
can detect the sharp cutoff in galactic emission in the far UV beyond
the 91.2 nm Lyman limit, the wavelength of light that can ionize
hydrogen, the most abundant element, and thus will be absorbed by it.
Galaxies at high redshifts have this Lyman break redshifted into
visible or even infrared light, where it can be observed from large,
ground-based telescopes.  A galaxy that is bright in colors with
wavelengths longer than the redshifted break, but disappears below it,
can be identified as being at a specific redshift \cite{steidel1996}.
Lyman break galaxies observed out to $z >8$ ($6\times10^8$~yr after
the Big Bang) can be used to determine the SFR density as a function
of redshift [e.g. \cite{bouwens2011}].  However, the faintest galaxies
at each redshift cannot be detected, so their contribution must be
extrapolated from the distribution of the brighter observed
galaxies. Only taking into account the directly observed galaxies
leads to the conclusion that the star formation rate drops off rather
quickly at high redshift \cite{kistler2009,moster2012,trenti2012} as
shown by the blue points in Fig.~1. However, this is in tension with
the first constraint, because it would imply a reionization redshift
somewhat later than the current concordance value $z = 10.1 \pm 1.0$
\cite{bennett2012}.
\begin{figure}
\begin{center}
 \includegraphics[width=0.7\textwidth]{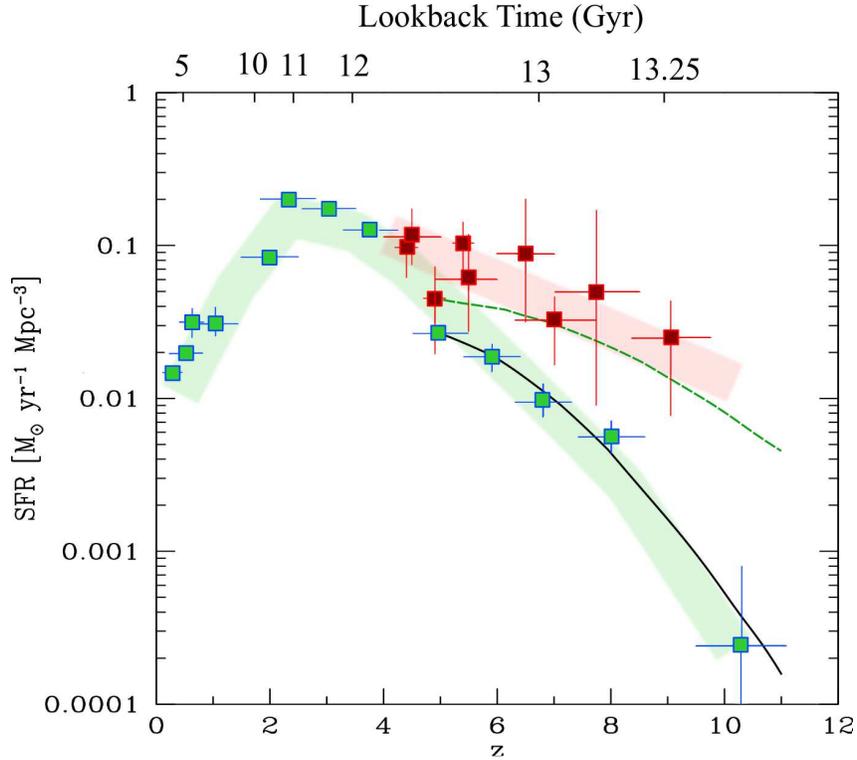}\end{center}
  \caption{{\bf Cosmic history of star formation.} The cosmic history of star formation in the universe versus
  redshift $z$ or lookback time \cite{trenti2012}. Star formation
  peaked some 10 billion years ago at a rate ten times as high as the
  modern era. At high redshift, the lower estimates outlined in green rely on
  observations of high-redshift Lyman break galaxies, neglecting those below the
  detection limit \cite{bouwens2011}, whereas the upper estimates
  outlined in red depend on the star formation intensity per gamma-ray burst
    \cite{kistler2009}.  The lines show model
    predictions \cite{trenti2010,robertson2012}, integrated over either the range of
    galaxies included in the Lyman break galaxy survey (black solid),
    or down to the expected lower limit for the mass of star forming
    galaxies (green dashed).
  }
\end{figure}

Gamma-ray bursts provide the second line of evidence.  These intense
flashes of tightly beamed, gamma-ray light produced in occasional
supernova explosions \cite{costa1997} are bright enough to be observed
back to $z > 8$. Derivation of the SFR associated with the host
galaxies of such high-redshift gamma-ray bursts depends on
calibrations at lower redshifts, though.  Extension of these
calibrations to high redshift requires determination of how the
incidence of gamma ray bursts as a function of the star formation rate
evolves with redshift \cite{kistler2009,robertson2012,trenti2012}.
One hypothesis is that gamma ray bursts occur primarily in galaxies
with low abundances of heavy elements, which form a greater
fraction of the population at high redshift.  However, the star
formation rate derived from that hypothesis (the red points in
Fig.~1) violates the second constraint, predicting too high a
stellar density by $z=4$ \cite{robertson2012}. Similarly, the model
shown by the green line in Figure~1, that includes this hypothesis, as
well as a physically-motivated minimum mass of dark matter halo in
which stars can form, still predicts slightly less star formation than
suggested by the red points. This suggests that some other effect must
also be acting \cite{robertson2012,trenti2012}.  The strength of this
effect can be empirically derived by measuring the fraction of gamma
ray burst host galaxies detectable at redshifts $z=5$ and $z =6$ \cite{trenti2012}.

\section*{Numerical Simulations}
Numerical simulations a decade ago [e.g. \cite{springel2003}] predicted that the peak in star
formation should occur at $z \sim 6$.  The contradiction to
the observations of the well-observed peak at $z \sim 2$ occurs in
large part because
neglecting feedback, or including
it by standard recipes \cite{cen1993}, leads to overproduction of stars
at early times. The alternative to date
has been reliance on ad hoc models of strong feedback, both from stars
and supermassive black holes, to suppress that
early star formation, in order to agree with the observations.

This is because simulations capturing cosmological scales have been unable to model
the interstellar medium within galaxies with sufficient resolution to
follow the energetics of stellar feedback successfully.  This leads to
the classical overcooling problem. Without local feedback that
transfers realistic amounts of energy to the interstellar medium, the
SFR can be an order of magnitude higher than observed, even in models
of modern galaxies \cite{katz1996,somerville1999,
  springel2003,keres2009,bournaud2010,
  dobbs2011,tasker2011,hummels2012}.

Another reason for requiring ad hoc models of energetic feedback,
which I will not focus on here, is that accretion into clusters of
galaxies produces huge reservoirs of hot gas.  The accretion of this
low density, hot gas onto massive elliptical galaxies must be
throttled. It is much easier to prevent accretion onto galaxies of low
density gas that can not radiatively cool easily, than it is to reheat
and expel already cooled and accreted gas. The required heating likely
comes from the jets driven by supermassive black holes in active
galactic nuclei, rather than stars.

Feedback models typically fail because of unphysical cooling in poorly
resolved regions of hot gas.
Interstellar and intergalactic gas cools radiatively
\cite{dalgarno1972,sutherland1993}: inelastic collisions excite
electrons into higher energy levels, while slowing down the colliding
particles (effectively reducing the temperature of the gas).  The excited electrons then drop
back to the ground state releasing photons.  So long as the densities
are low enough for collisions not to deexcite the electrons prior to
radiation, and the opacity of the gas and dust is low enough to allow escape of
the photons from the system, these inelastic collisions lead to energy loss, and thus
cooling. Interstellar and intergalactic gas away from the very densest
cores of star forming regions typically satisfies both of these
conditions.  

Because radiative cooling relies on collisions, its strength $\dot{E} = - n^2
\Lambda(T)$ depends on the square of the number density $n$, as well
as having a strong temperature dependence $\Lambda(T)$ from the
distribution of available energy levels for electron excitation.
Crucially, the value of $\Lambda(T)$ for $T = 10^5$~K gas in
ionization equilibrium exceeds by more than an order of magnitude that
for hot $10^6$~K or cool $10^4$~K gas  [Fig.~2a, \cite{sutherland1993}].  The
elevated cooling around $10^5$~K occurs because of the ability of
collisions in that temperature range to excite the strong resonance
lines of lithium-like ions of carbon, oxygen, and nitrogen, the most
common elements heavier than helium.
\begin{figure}
\begin{center}
\includegraphics[width=0.7\textwidth]{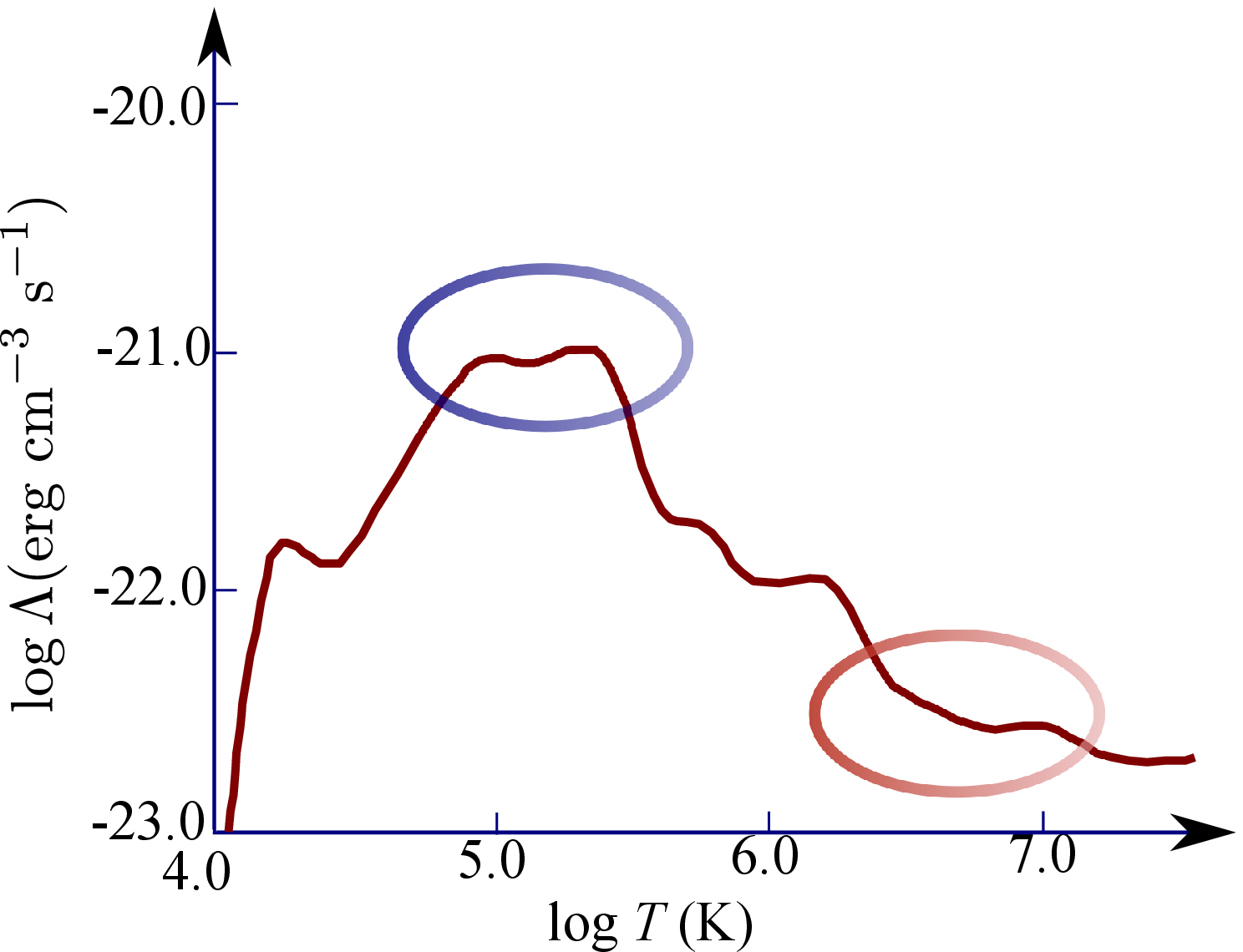}
 \includegraphics[width=0.7\textwidth]{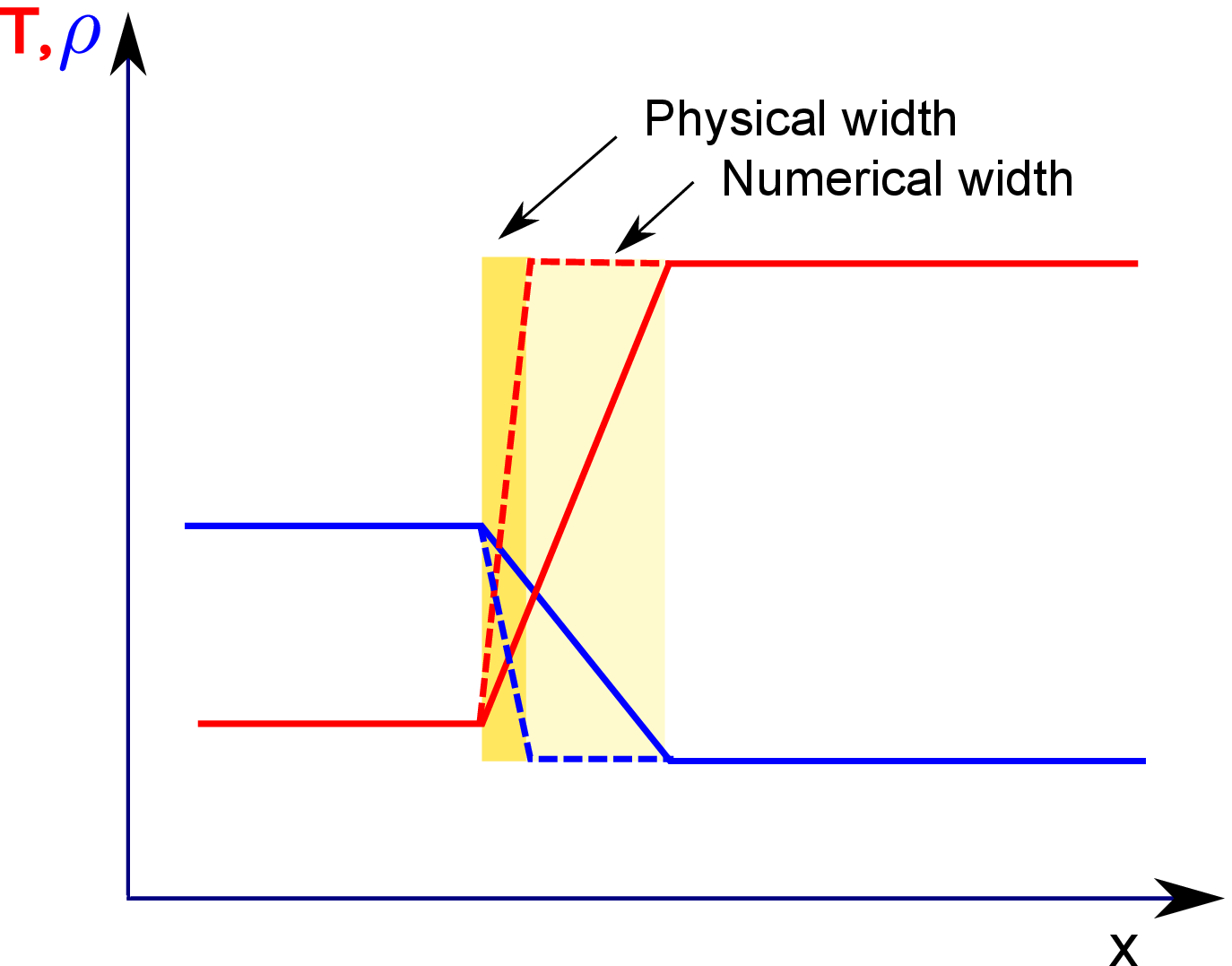}\end{center}
\caption{{\bf Cooling of interstellar gas.} (a) The temperature dependence of the cooling function
  $\Lambda(T)$ from \cite{sutherland1993}, showing the order of
  magnitude 
  higher cooling rate at around $10^5$~K (red circle) compared to $10^{6-7}$~K
  (blue circle).  (b) Sketch of how unphysical
  cooling can occur at poorly resolved numerical interfaces when too
  much intermediate temperature gas is present.}
\end{figure}

The temperature and density dependence of the cooling function leads
to two serious problems for numerical simulations. First, supernova
explosions drive blast waves that shock gas to temperatures above 
$10^6$~K, so that it only cools with difficulty. However, if stellar
feedback energy is fed into the grid of a numerical simulation too
slowly, or over too large a volume, it will only raise the temperature
into the $10^5$~K range or lower, so that the energy will promptly radiate
without exerting dynamical effects.  Second, simulations
of gas flow usually require several zones to resolve interfaces
between gas with different properties, such as hot and rarefied or
cold and dense. Within such an interface, the resolution of the
numerical grid determines the amount of intermediate
temperature gas, rather than the physics of the interface (See Fig.~2b).  In poorly
resolved models, such interfaces capture large volumes of gas 
that cools unphysically.  Over 25 years ago, Tomisaka
\cite{tomisaka1986} already demonstrated that the evolution of superbubbles
formed by multiple supernova explosions from an association of OB
stars could not be adequately simulated with 5~pc (16 light year) resolution because
of such strong numerical overcooling \cite{mac-low1989}.
For models of the diffuse interstellar medium of the Milky Way galaxy
($0.01 < n < 100$~cm$^{-3}$), it has been demonstrated that 2~pc (6.5
light year) grid cells resolve interfaces well enough to avoid
dynamically important loss of energy from hot gas
\cite{avillez2000,joung2006,hill2012}. Such a model, from
\cite{hill2012} is shown in Movie S1. 

\section*{How Does Stellar Feedback Inhibit Star Formation?}

Energetic feedback from stars heats the interstellar gas through radiative
ionization, and drives strong shock waves through it from supernova
explosions, leaving it in a characteristic state of supersonic, highly
compressible turbulence.  This differs in its properties from the
almost incompressible turbulence familiar from the terrestrial
atmosphere and ocean because of the strong density fluctuations
produced by the shock waves, and the strong vorticity sheets produced
at shock intersections.

Such highly compressible turbulence both promotes and prevents
gravitational collapse.  We can heuristically estimate which effect wins by
examining the dependence of the Jeans critical mass for gravitational
collapse (eq.~1) $M_J \propto \rho^{-1/2} c_s^3$ on the
root-mean-square turbulent velocity $v_{\rm rms}$
\cite{mac-low2004}. In the classical picture, turbulence is treated as
an additional pressure \cite{chandrasekhar1951,von-weizsaecker1951},
so that we can define an effective sound speed $c_{s,{\rm eff}}^2 =
c_s^2 + v_{\rm rms}^2 / 3$. This additional effective pressure
increases the Jeans mass by $M_{\rm J} \propto v_{\rm rms}^3$,
inhibiting collapse.  On the other hand, shock waves with Mach number
${\cal M} = v_s / c_s$ in an isothermal medium cause density
enhancements with density ratio $\rho_s / \rho_0 = {\cal M}^2$.  These
turbulent compressions decrease the Jeans mass by $M_J \propto
\rho_s^{-1/2}$, assuming that the shocks typically have shock velocity
$v_s \simeq v_{\rm rms}$.

When we combine the effects of increased pressure and increased
density, we find that
\begin{equation}
M_{\rm J} \propto \left(\frac{c_s}{v_{\rm rms}}\right) \left(c_s^2 +
  \frac{v_{\rm rms}^2}{3}\right)^{3/2} \propto v_{\rm rms}^2
\end{equation}
for $v_{\rm rms} \gg c_{\rm s}$. Thus, highly supersonic turbulence
strongly inhibits collapse.  Because turbulence acts intermittently,
however, it still promotes collapse locally, in shock compressed
regions, even though its net effect is to inhibit collapse globally.
Therefore, a region with mass lower than the Jeans mass globally
because of turbulence can still have isolated regions of local
collapse, as shown in isothermal simulations \cite{klessen2000}. The
role of compression was isolated in simulations by
\cite{federrath2013} who increased the average density with the
dependence on the Mach number required to balance the turbulent
support term.  Even higher resolution simulations using adaptive mesh
refinement \cite{padoan2012} show that the star formation efficiency
can be expressed in terms of the ratio of the crossing time $t_x = L/2
v_{\rm rms}$ for a region of size $L$ to the free-fall time $t_{\rm
  ff}$.

The net suppression of star formation by turbulence has
consequences in the diffuse, stratified interstellar medium of
galaxies. This was demonstrated \cite{joung2006} with a well-resolved model of
supernova driving of turbulence, including a treatment of radiative
heating and cooling, but not self-gravity of the gas, run with the
Flash adaptive mesh refinement gas dynamics simulation code
\cite{fryxell2000}. The turbulent flow indeed compressed gas into
Jeans-unstable regions of cold, dense gas with sizes comparable to
observed star-forming clouds.  However, if the SFR
expected for those regions is computed, it is an order of magnitude
less than the rate required to produce the assumed number of
supernovas driving the trubulence.  This is consistent with the
results of uniformly-driven, self-gravitating turbulence described in
the preceding paragraph.  Turbulence triggers star formation
inefficiently; it does not lead to
stochastic propagating waves of star formation as once envisioned by
\cite{gerola1978}. 

At the scale of single star-forming clouds, the question of triggering
has also been studied \cite{dale2007}. Here the most important effect
is heating of the gas from under 100~K to around $10^4$~K when UV
radiation from newly formed massive stars ionizes it
\cite{elmegreen1977}. The resulting huge increase in pressure drives a
blast wave outwards. This strongly disturbs the morphology of the gas
cloud.  However, the actual difference in the SFR is
small, accelerating the formation of stars by only 20\% of the
free-fall time of the cloud.

Quantitative observational studies support the conclusion that
triggering does not represent a major mode of star formation.
Although triggered star formation clearly occurs around regions of
massive star formation, it is a relatively small effect that does not
explain most star formation, consistent with the 10\% effectiveness
found by \cite{joung2006}. For example, even under favorable
circumstances, compression of gas by nearby massive stars triggers
less than a quarter of star formation in the Elephant Trunk Nebula
\cite{getman2012}. At the galactic scale, multiple supernova
explosions sweep up supershells in the Large Magellanic Cloud. Only
12--25\% of the star-forming gas traced by the emission of the
molecule carbon monoxide in the supershells formed as a direct result
of supershell formation, corresponding to no more than 11\% of the
total star-forming gas in this galaxy \cite{dawson2013}.

\section*{What are the Sources of the Interstellar Turbulence?}

If turbulence limits star formation, then understanding sources of
turbulence, both in the diffuse gas and in dense clouds of
star-forming, molecular gas, will help us to understand star formation.

Supernova explosions effectively drive turbulence in the diffuse gas
\cite{mac-low2004}. A study \cite{joung2009} of the velocity
dispersion resulting from supernova rates ranging from the Milky Way
value to 512 times higher, as would be seen in extreme starburst
galaxies having SFRs hundreds of times that of the
Milky Way, showed that, regardless of the supernova rate,
supernova driving resulted in a rather uniform velocity dispersion
$v_{\rm rms} = 5$--10~km~s$^{-1}$. This work varied the surface
density of the gas disks following the Kennicutt-Schmidt law
\cite{kennicutt1998} relating the surface density to the SFR, and thus
the supernova rate.
To simulate observations in the 21~cm line of atomic hydrogen, they
fit single Gaussian components 
to the velocity of gas in the atomic temperature range
(roughly $10^2$--$10^4$~K).  The resulting simulated 
observations showed
spectral line widths equivalent to 10--20~km~s$^{-1}$, agreeing with
most observations \cite{petric2007,tamburro2009}, aside from extreme
starbursts where elevated 21~cm line widths are observed
\cite{murray2010}.

The driving of interstellar turbulence in nearby galaxies was more
generally studied by \cite{tamburro2009} in a sample of galaxies
observed with the Spitzer Infrared Nearby Galaxies Survey \cite{SINGS}
and The H~{\sc i} Nearby Galaxies Survey \cite{THINGS}.  The energy
input from supernovas can explain the observed density of kinetic
energy of the interstellar gas within the star-forming inner regions
of the observed disk galaxies. However, in the primarily gaseous outer disks, where star
formation drops off strongly, so there are few supernova explosions,
some other mechanism has to be stirring 
the gas to maintain the observed density of kinetic energy.

Two possibilities have been proposed for this alternative
mechanism. One possibility is magnetorotational instability
   \cite{balbus1998},
which transfers energy from differential rotation into turbulence 
whenever the angular velocity decreases
outward in rotating, magnetized gas. Galactic disks have flat rotation curves, with
constant orbital velocity, because they lie within massive haloes of
dark matter.  Therefore their angular velocity indeed decreases
outward, so magnetorotational instability can drive substantial
turbulence \cite{sellwood1999}.

The strength of this turbulence depends, however, on the
thermal properties of the gas.  The temperature dependence of the
cooling curve for interstellar gas at $T < 10^4$~K leads to the
possibility of a two-phase medium, in which the gas has two stable
equilibria with the radiative heating \cite{field1965}, one 
cold ($T \sim 10^2$~K) and dense ($n \sim 100$~cm$^{-3}$) phase, and one warm
($T \sim 10^4$) and lower density ($n \sim 1$~cm$^{-3}$) phase.
Simulations have shown that velocity dispersions
of the magnitudes observed in 21~cm emission can be reached if such a
two-phase medium forms 
\cite{piontek2007}.  The energy input from the magnetorotational
instability appears sufficient to explain the kinetic energy seen in
the outer disks of galaxies beyond the star-forming region where supernovas are expected 
\cite{tamburro2009}.  

Another possibility, though, is that radiative heating from external
sources of UV radiation such as quasars or starburst galaxies can
by itself maintain the observed velocity
dispersion as thermal motion \cite{elmegreen1994,schaye2004}.  Under
this hypothesis, gas in the outer disk must lie predominantly in the warm
phase, and the transition to a two-phase medium as density increases
marks the radius at which star formation begins. This model can be
distinguished observationally from magnetorotational instability by
its prediction of an absence of gas in the cold, dense phase in outer
disks.  The discovery of finite rates of star formation in outer disks
by the GALEX satellite UV observatory \cite{boissier2007}, however,
suggests the presence of the cold, dense phase, supporting
magnetorotational instability as the second driving mechanism.

Recently, radiation pressure from the light emitted by the most
massive star clusters reflecting off of dust grains in the
interstellar gas has been argued to play a major, or even dominant,
role in limiting star formation in galaxies
   \cite{murray2010}.
However, this
conclusion depends on how well radiation can couple to the dust, and
thence with the gas by collisions of moving dust grains with the
surrounding gas particles.  If each photon from the massive stars only
scatters once off of a dust grain before escaping the system, then the
radiation pressure from a star cluster with luminosity $L$ is
proportional to $L/c$, where $c = 3 \times 10^5$~km~s$^{-1}$ is the
speed of light.  This is sometimes called the momentum-driven limit,
as this conserves the momentum of the radiation, but much of its
energy escapes.  If, on the other hand, the dust is extremely
optically thick, so that the photons continue scattering off of the
dust until they lose almost all their energy, then the radiation
pressure is far higher, proportional to $L/v_{\rm rms}$, called the
energy-driven limit.  
Because the turbulent motions in interstellar gas
have $v_{\rm rms} \sim 10$~km~s$^{-1}$, this represents a
huge difference.

Although it is unlikely that the energy-driven limit is ever reached
in real star-forming galaxies, the argument for the importance of
radiation pressure relies on the expectation that the number of times
the photons scatter is comparable to the infrared optical depth
$\tau_{\rm IR}$ of the most massive star-forming regions in the
galaxies, which can be over 100.  This assumption suggests a radiation
pressure proportional to $\tau_{\rm IR} L / c$.  Several groups
\cite{krumholz2009,fall2010} argued instead for the momentum-driven
limit actually being the appropriate one, however, suggesting that
radiation pressure is far less important in galactic evolution.

Recent multi-dimensional simulations of radiation pressure acting on a
layer of gas with optical depth $\tau_{\rm IR} \gg 1$
\cite{krumholz2012a} showed that the
radiation acts as a rarefied fluid accelerating a dense fluid, which
causes Rayleigh-Taylor instability.  The instability overturns and
fragments the dense gas, indeed stirring it, but allowing the
radiation to escape far more quickly than would be expected from its
initial optical depth. 
As a result
of this overturn, although radiation pressure indeed is more effective
than the momentum-driven limit, it is typically at least an order of
magnitude less efficient than suggested by the assumption of
proportionality to $\tau_{\rm IR} L / c$.  This calls into serious
question results based on that assumption.

The gravitational interaction of the gas with itself can temporarily drive
turbulence even in the absence of outside energy inputs.  Such gravitationally-driven
turbulence indeed occurs in disks that are dense and cool enough to be subject to
gravitational instability \cite{bournaud2010}. The resulting
turbulence produces a distribution of surface density fluctuations
consistent with observations of atomic gas in the Magellanic Clouds.
However, turbulence decays in roughly a gravitational free-fall time
  \cite{mac-low1999a}
so turbulence driven by
internal gravitational motions cannot effectively delay gravitational
collapse and subsequent star formation on its own.

On the other hand, accretion of gas from the intergalactic medium onto
galaxies, a process that appears to continue to the present
day, carries substantial kinetic energy with it, so it could feed
gravitationally-driven turbulence.  
Even if that energy
only couples to the interstellar gas with 10\% efficiency, the
velocity dispersion observed in the gas of galaxies comparable in mass
to the Milky Way could be maintained if they accrete gas at the same
rate as they form stars \cite{klessen2010}.  Lower mass dwarf galaxies have lower
accretion velocities, though, and so gain less energy.  Nevertheless,
they have the same observed velocity dispersion, which cannot be
explained by this mechanism.  Such dwarf galaxies represent the
dominant location for star formation over cosmic time
\cite{karim2011}, so other mechanisms than accretion, such as
supernova driving, must play an important role in its regulation.

    Accretion may, however, indeed dominate the
dynamics of individual star-forming clouds.  They appear to
continually accrete gas from the surrounding interstellar medium at a
rate sufficient to drive the turbulent motions of 1--5~km~s$^{-1}$ observed within them
\cite{klessen2010,vazquez-semadeni2010,goldbaum2011}.
This results in longer
lifetimes than would be expected for isolated clouds of the same mass,
because the driven turbulence inhibits collapse and star
formation. However, other forms of feedback such as ionization heating
appear necessary to explain how star formation within these clouds
comes to an end and the dense gas disperses.

Models of galactic evolution do lead us to one firm conclusion:
driving of turbulence by internal gravity or accretion onto galaxies
must be supplemented by other energy sources, as \cite{klessen2010}
argued must be true at least for dwarf galaxies.  Otherwise, standard
simulations performed without stellar feedback or other energy sources
beyond gravity would be sufficient to reproduce observed galaxies,
which is not the case, as summarized by \cite{hummels2012}, and
simulated at high resolution in small, isolated dwarfs by
\cite{simpson2012}. Insufficient numerical resolution leading to
excessive dissipation of turbulent energy may play a role
in massive galaxies, but not in the well-resolved dwarf simulations.
Ultimately, gravity must compete with stellar feedback and other
energy sources to determine the progression of collapse and star
formation.

\section*{Star Formation Laws}
Star formation correlates well with gas surface density in galaxies.
The most well-known version of this correlation, the Kennicutt-Schmidt
law \cite{kennicutt1998},
relates the SFR surface density $\Sigma_{\rm SFR}$ to
the total (both atomic and molecular) gas surface density $\Sigma_{\rm
  gas}$ averaged over either the entire disks of
normal or starburst galaxies, or entire galactic centers to derive
the empirical law 
\begin{equation}\Sigma_{\rm SFR} \propto \Sigma_{\rm
    gas}^{1.4}. \end{equation}
Although variations around this law have been repeatedly found, for
example in \cite{shetty2013}, it nevertheless appears to broadly hold
\cite{kennicutt2012}. 
As observations reached higher resolution and were able to resolve
regions a few thousand light years on a side in nearby galaxies, a
similar correlation was found for them \cite{bigiel2008}.  However, in
this case (Fig.\ 3a), regions with the lowest gas surface
densities have lower SFR densities than would be expected from the
power-law correlation, whereas the highest gas surface densities have
higher than expected rates. Comparison to nearby star forming regions
in the Milky Way shows that individual molecular clouds dozens of
light years in size form stars far more efficiently than the larger
regions observable in external galaxies \cite{heiderman2010}. The
rates observed in local clouds have been successfully simulated using
high-resolution simulations of super-Alfv\'enic, isothermal
turbulence \cite{federrath2012}.
\begin{figure}
 \begin{center}
  \includegraphics[width=3in]{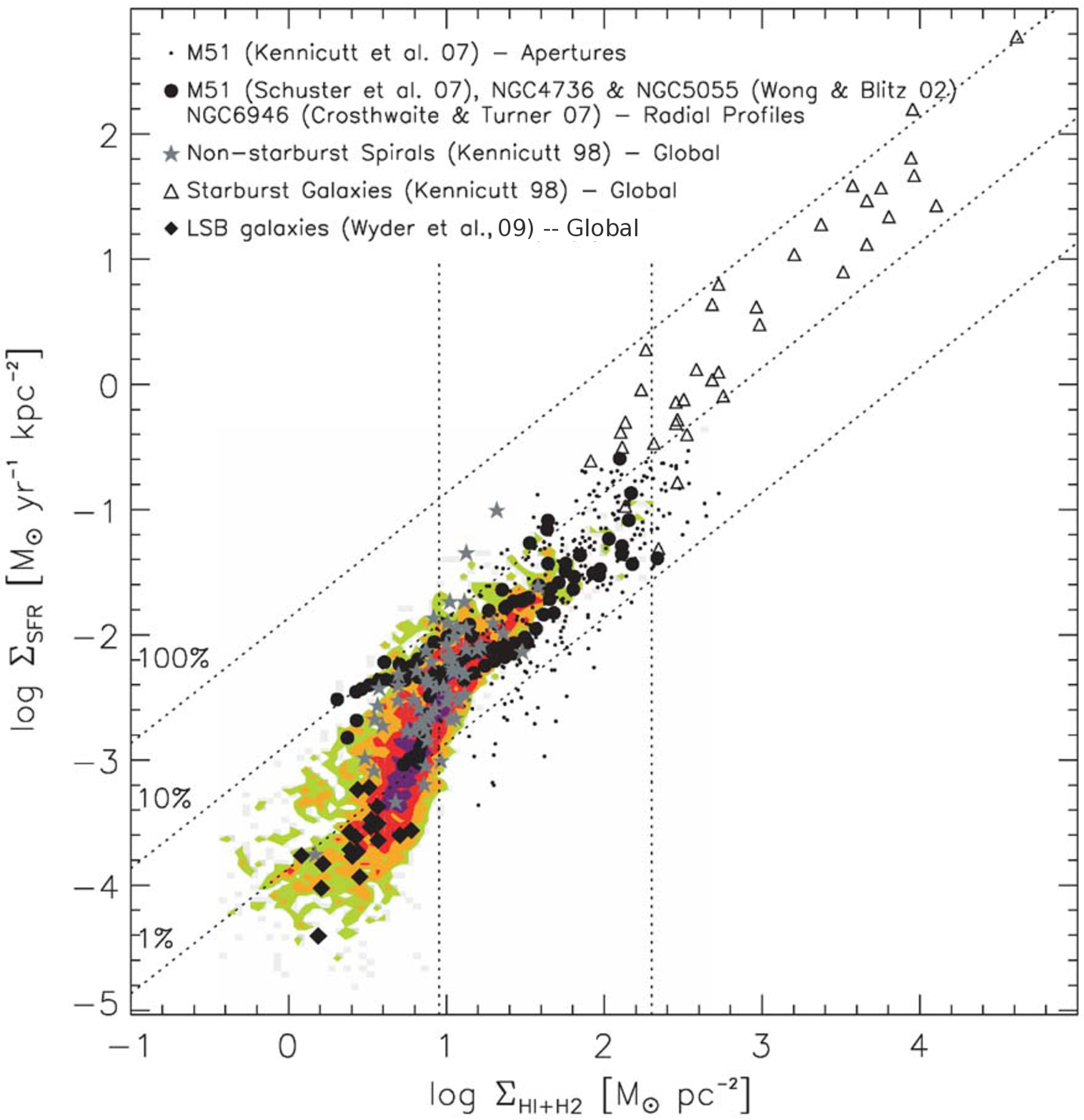}
   \includegraphics[width=3in]{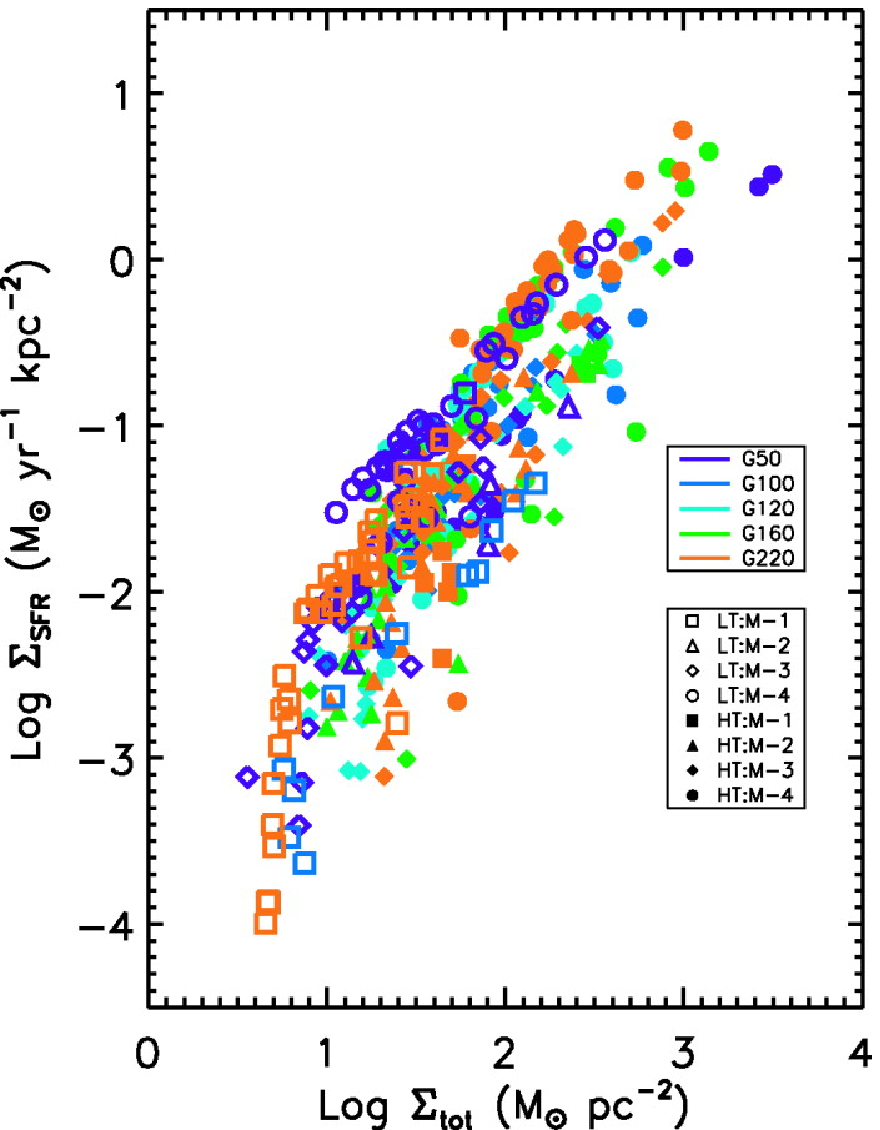}
  \end{center}
\caption{{\bf Star formation correlations.}  {\em (a)} A comparison of SFR surface density $\Sigma_{\rm
    SFR}$ to total gas surface density
  $\Sigma_{\rm gas}$ from observations presented in \cite{bigiel2008} showing
  the combined data from that paper in colored contours, along with points
  from the observations described in the legend on the figure\cite{kennicutt2007,schuster2007,wong2002,crosthwaite2007,kennicutt1998,wyder2009}.  The
  dashed lines show what percentage of the gas would be consumed at that
  SFR over a period of $10^8$~yr. {\em (b)} Radial
  profiles across model disks simulated with isothermal gas and live
  stellar disks and dark matter halos \cite{li2005a}, showing the
  same drop in star formation efficiency at low gas surface density.}
\end{figure}

\framebox{\begin{minipage}{6in}\paragraph{Star formation rate} Observers derive SFRs from
different observational indicators that trace either directly or
indirectly the ionizing UV radiation from massive stars.  Such stars have
short lifetimes of only a few million years, so their presence traces
recent star formation.  The dense gas and dust immediately around
newly forming stars absorbs UV light efficiently, ionizing and heating up in the
process and reemitting the energy in the far infrared. Eventually the
expanding bubble of ionized gas escapes the cloud.  That gas can be
observed in the H$\alpha$ line (the 2-1 transition of recombining
atomic hydrogen), which lies in the red portion of the visible
spectrum.  After the young stars have entirely escaped their natal
gas, they can be directly observed in the UV.

Determination of the total gas surface density depends on observation
of not just the atomic hydrogen that dominates the diffuse
interstellar gas, but also the H$_2$ that forms in the
dense, star-forming clouds.  However, the lowest rotational lines from
the low-mass H$_2$ molecule lie at such high energies that they are
not excited at the low temperatures typical of molecular gas (e.g. \cite{kennicutt2012}).
Instead, emission from the next most common
molecule, carbon monoxide, is used as a tracer of the
molecular gas.  This requires the use of a conversion factor $X_{\rm
  CO}$ that has been found to be roughly constant in the Milky Way.
However, it has recently been demonstrated
\cite{glover2011,narayanan2011,schruba2011,shetty2011a} that the
conversion factor rises sharply in regions with low absorption due to
either low column density of gas and dust, or low heavy element
abundance leading to low dust fractions in the gas.  Conversely, it
seems to drop
in high column density regions dominated by molecular gas
\cite{narayanan2012}, such as in starburst galaxies \cite{downes1998}.
\end{minipage}
} 

While star formation deviates from the power law correlations with
total gas in extreme cases, it correlates linearly with H$_2$ surface
density over the entire range of observed values
\cite{rownd1999,wong2002,bigiel2008,leroy2008,bigiel2011}, although
with more than an order of magnitude scatter among individual regions.
This has been interpreted to mean that H$_2$ formation causes or
controls star formation \cite{robertson2008,krumholz2009,gnedin2010,
christensen2012}. However,
one must ask whether correlation actually implies causation.

Indeed other measurements of high density gas (($n>10^4$~cm$^{-3}$)
also show linear correlations of the column density of the gas with
star formation.  These include a linear correlation between HCN
emission, which only occurs above a critical density $n_c \sim
10^5$~cm$^{-3}$, and $\Sigma_{\rm SFR}$ \cite{gao2004}; and a direct
correlation between the number of young stellar objects in a region,
and the mass of material with infrared extinction in the 2~$\mu$m K-band $A_K >
0.8$ \cite{lada2010} magnitude.

Examination of the physics of star formation reveals that molecule
formation is not vital to the process of gravitational collapse, so
long as heavy elements such as carbon and oxygen are present even in
trace quantities $> 10^{-5}$ the solar abundance.  It had been argued
that low-energy molecular lines were required to allow radiative
cooling of star-forming gas to only a few tens of degrees above absolute
zero.  Molecular gas does, in fact, dominate the cooling of such high
density gas, but this is coincidental: pure atomic gas at the same
densities cools almost as effectively by radiation from fine structure
lines of heavy elements.  Instead, the key factor for star formation
seems to be shielding of the cold gas by dust from heating by
background UV light \cite{krumholz2011,glover2012}.  Removing
molecular cooling from models changes the minimum temperature from 5~K
to 7~K, whereas removing the shielding increases the minimum
temperature by more than an order of magnitude \cite{glover2012}.
Indeed, well-resolved galaxy formation simulations reached virtually the same result
whether star formation was limited to occur only in molecular gas, or
allowed to occur in all dense gas \cite{hopkins2011}.

The correlation observed between H$_2$ and star formation rate surface
densities appears not to be causal, but rather to occur because both
have a common cause.  Formation of H$_2$ occurs over a timescale
\cite{hollenbach1971} of $t_f = (1 \mbox{ Gyr}) / (n / 1 \mbox{
  cm}^{-3})$.  As stars form through gravitational collapse, high
densities are inevitably reached, and H$_2$ then forms quickly
\cite{glover2007}.  This happens almost independent of the dust-to-gas
ratio \cite{glover2012a}, which is largely controlled by the abundance
of heavy elements, even though H$_2$ formation depends on dust
surfaces.  The dust grains provide the necessary catalyst for
formation of a homonuclear molecule at densities low enough that
three-body collisions hardly ever occur. (An exception to the need for
dust surfaces occurs in the almost radiation-free environment of the
early universe where electrons and protons can act as catalysts in gas
phase reactions involving the fragile H$^-$ and H$_3^+$ ions).

Collapse occurs within a free-fall time \cite{krumholz2012}.  However,
at solar elemental abundances, cooling occurs far more quickly, with
the cooling time $t_{\rm cool} < t_{\rm ff}$ for abundances as low as
$10^{-4}$ of the solar abundance, characteristic of the early universe
or the very lowest abundance dwarf galaxies in the local
neighborhood. In such low abundance gas, molecule formation is delayed
severely compared to the near solar abundance gas characteristic of the
Milky Way.  Then H$_2$ only forms in the very densest cores of the
collapsing region, leading to low integrated molecular fractions,
despite ongoing star formation.

\section*{Gravitational Instability}

The gravitational instability of the gas and stars in galaxies does
appear to control star formation directly.  We can heuristically
derive \cite{schaye2004,mac-low2004} the criterion for stability of a differentially rotating
galactic disk \cite{toomre1964,goldreich1965} by comparing the
free-fall time required for collapse of 
a region to the time required for it to shear apart, or for a pressure
wave to cross it.  We consider a thin galactic disk with uniform
sound speed $c_s$ and surface density $\Sigma$.  The
Jeans criterion for gravitational instability requires that the time
scale for collapse of a density perturbation of size $\lambda$ 
\begin{equation}
t_{\rm ff} = (\lambda / G \Sigma)^{1/2}
\end{equation} 
not exceed the time required for the gas to respond to the collapse by
changing its pressure, the sound crossing time
\begin{equation}
t_{\rm s} = \lambda / c_s.
\end{equation}
This implies that regions can collapse if they have size
\begin{equation}
\lambda > c_s^2 / G \Sigma.
\end{equation}
In a 
rotating disk, 
collapsing perturbations 
effectively rotate around themselves because of Coriolis forces \cite{toomre1964,elmegreen2011} ,
causing centrifugal motions that can
also support against gravitational collapse if the collapse time scale
$t_{\rm ff}$ exceeds the rotational period $t_{\rm rot} = 2 \pi
\kappa$, where the epicyclic frequency $\kappa$ is of order the
rotational frequency of the disk $\Omega$. Thus, for collapse to
proceed,
\begin{equation}
\lambda < 4 \pi^2 G \Sigma / \kappa^2.
\end{equation}
Gravitational instability occurs if there are regions with sizes that satisfy
both of these criteria simultaneously, having
\begin{equation}
\frac{c_{\rm s}^2}{G\Sigma} < \lambda < \frac{4 \pi^2 G\Sigma}{\kappa^2}.
\end{equation}
This occurs if the Toomre parameter \cite{toomre1964} 
\begin{equation} 
Q \equiv c_s \kappa / (2 \pi G \Sigma) < 1.
\end{equation} 
The full criterion for instability derived from linear analysis of the
equations of motion of gas in shearing disks gives a factor of $\pi$
rather than $2\pi$ in the denominator
\cite{safronov1960,goldreich1965}, whereas using kinetic theory 
for collisionless stellar disks gives a factor of 3.36
\cite{toomre1964}. 

When collisionless stars and collisional gas both contribute to
gravitational instability, as in most galactic disks, a rather more complicated formalism is
required to accurately capture their combined action
\cite{gammie1992,rafikov2001}.  This has been
successfully approximated with simple algebraic combinations of the stellar and
gas Toomre parameters \cite{wang1994,romeo2011}.  Gas supported by
turbulent flows rather tan pure thermal pressure has no formal
minimum wavelength for gravitational collapse in the presence of
turbulent dissipation \cite{elmegreen2011},
although finite disk thickness does act to stabilize the smallest
wavelengths against collapse.

Numerical simulations display the relationship between global
gravitational instability and star formation.  For example,
\cite{li2006} simulated the behavior of exponential gas disks embedded
in live stellar disks, with the turbulent velocity in the disk modeled
using an isothermal equation of state having a 6--12~km~s$^{-1}$ sound
speed.  They varied the strength of the gravitational instability, and
measured how much gas collapsed gravitationally in models that
resolved the Jeans length to densities of $n \sim 10^7$~cm$^{-3}$.  They found not only
that all of their models fell cleanly on the global Kennicutt-Schmidt correlation
between total gas and SFR surface density, but also (Fig.~3b), that
analysis of azimuthal rings in their models 
predicts the falloff in star formation at low gas densities observed
at kiloparsec scales \cite{bigiel2008}. 

Another example of the strength of the hypothesis that gravitational
instability controls star formation lies in the unusual morphologies
of many high-redshift galaxies. Clumpy, irregular galaxies occur far
more frequently at $z \sim 2$ than in modern times
\cite{elmegreen2009}.  Galaxies then tended to be far more gas-rich
than now because gas accreted far more quickly in the denser
high-redshift universe.  As a result, high-redshift galaxies are more
likely to be strongly gravitationally unstable.  Well-resolved
adaptive-mesh computations show that such conditions naturally lead to
the formation of giant, gravitationally bound clumps \cite{agertz2009}
consistent with the observed morphologies.

An extended version of the gravitational instability hypothesis
\cite{ostriker2010,ostriker2011,kim2011,shetty2012} proposes that star
formation is controlled by the combination of gravitational
instability and thermal equilibrium in a two-phase medium.  This
occurs because stellar feedback increases as star formation increases,
which in turn is driven by gravitational instability.  However, as the
feedback increases, it heats and stirs the gas.  This reduces its
gravitational instability as its effective pressure increases, and
ultimately prevents gravitational instability entirely if the pressure
increases beyond the range available to a two-phase medium.  Thus, in
this model, the star formation rate adjusts until a steady-state rate
of star formation is reached.

\section*{Outlook}

Understanding the cosmic history of star formation requires a statistical
description of the luminosity function of galaxies, particularly of faint galaxies
at early times. 
These faint galaxies dominate cosmic reionization \cite{bouwens2012}
and can explain most of the apparent difference between star formation
histories derived from gamma ray burst statistics and Lyman break
galaxies \cite{trenti2012}.  Heroic efforts with the final instrument
set on the {\em Hubble Space Telescope} are yielding our first
glimpses into this territory \cite{trenti2011,bouwens2012}, but real
progress will occur with the next generation of ground and space based
telescopes such as the {\em James Webb Space Telescope}
\cite{gardner2006}, the European Extremely Large Telescope
\cite{lyubenova2009} and other 30 meter telescopes, and the Large Synoptic Survey Telescope
\cite{tyson2002}.

Resolving feedback in simulations reaching cosmological scales remains
tremendously difficult, but computers are slowly becoming sufficiently
powerful and algorithms well developed enough for this to fall within
the realm of the possible.  Simulations of star formation over cosmic
time that result in realistic star formation histories for single
galaxies \cite{hopkins2011,governato2012} and even clusters of
galaxies \cite{cen2011a} have started appearing.  Expanding these to
statistically representative samples of the universe containing both
clusters and voids remains a huge challenge that is just starting to
be met \cite{vogelsberger2012}. Use of adaptive spatial resolution on either
structured \cite{kravtsov1997,o-shea2004} or unstructured
\cite{springel2010,vogelsberger2012} meshes\footnote{see
http://www.cfa.harvard.edu/itc/research/movingmeshcosmology/ for visualizations
of galaxies from unstructured mesh simulations} will clearly play a central role in making
progress. Also required will be use of hybrid algorithms to take full
advantage of massively parallel computer clusters made up of nodes
with multiple processors sharing memory or even graphics coprocessors.

As our understanding of star formation comes into focus, we will be
able to apply the knowledge gained to a series of outstanding
questions. One is the evolution of heavy element abundances in
galaxies, ultimately leading to planet formation around stars with
sufficient rock-forming elements \cite{buchhave2012}.  Another is
understanding the structure and origins of our own Milky Way galaxy,
while a third is understanding how gas between galaxies both in
clusters and in the field gets polluted with heavy elements and heated
up to observed abundances and temperatures. Finally, the very
structure of the cores of dark matter halos appears intertwined with
star formation and feedback.  Thus, improving our understanding of
star formation will provide the key to unlocking the story of our own
origins and those of the universe around us.

\bibliography{1229229bibfile}

\bibliographystyle{Science}


\begin{scilastnote}
\item An early version of this review was presented at IAU
  Symposium 292 \cite{mac-low2013}. I have benefited over the past 15
  years from discussions and collaborations on these topics with
  M. A. Avillez, B. G. Elmegreen, C. Federrath, S. C. O. Glover,
  F. Heitsch, A. S. Hill, M. R. Joung, R. S. Klessen, M. R. Krumholz,
  Yuexing Li, T. Peters, and E. V\'azquez-Semadeni.  I thank
  J. Beacom, C. Federrath, S. C. O. Glover, and J. C. Iba\~nez for useful comments on
  the manuscript, and the anonymous referees for detailed and
  constructive reviews that improved this work. I was partially supported by NSF grant
  AST11-09395, NASA Chandra Theory grant TMO-11008X, and DFG
  Sonderforschungsbereich 881---The Milky Way System.

\end{scilastnote}






\end{document}